# Quantum Oscillations in Hole-Doped Cuprates

Suchitra E. Sebastian,[1] Cyril Proust[2]

[1]Cavendish Laboratory, Cambridge University, Cambridge CB3 OHE, U.K.,
[2]Laboratoire National des Champs Magnétiques Intenses, Toulouse 31400, France

**One of the leading challenges of condensed matter physics in the past few decades in an understanding of the high-temperature copper-oxide superconductors. While the $d$-wave character of the superconducting state is well understood, the normal state in the underdoped regime has eluded understanding. Here we review the past few years of quantum oscillation measurements performed in the underdoped cuprates that have culminated in an understanding of the normal ground state of these materials. A nodal electron pocket created by charge order is found to characterise the normal ground state in $YBa_2Cu_3O_{6+\delta}$ and is likely universal to a majority of the cuprate superconductors. An open question remains regarding the origin of the suppression of the antinodal density of states at the Fermi energy in the underdoped normal state, either from mainly charge correlations, or more likely, from mainly pairing and / or magnetic correlations that precede charge order.**

## 1 HIGH TEMPERATURE SUPERCONDUCTIVITY

Superconductivity is the extraordinary phenomenon by which certain special materials lose all resistance to the flow of electricity below a certain 'superconducting' temperature. The phenomenon of superconductivity was first discovered at 4.2 K in mercury in 1911. Since then, the superconducting mechanism and maximum achievable superconducting temperature in similar families of conventional superconducting materials has been well understood. This is known as the BCS (Bardeen-Cooper-Shrieffer) theory,[1] which postulated that superconductivity is an electronic instability of the normal state, e.g. of the Fermi liquid state for conventional metal. It was therefore a revelation when, more than 75 years after superconductivity was first discovered, a new family of copper-oxide superconductors was discovered that shattered both the previous ceiling on superconducting temperatures and the previous understanding of the superconducting mechanism.[2]

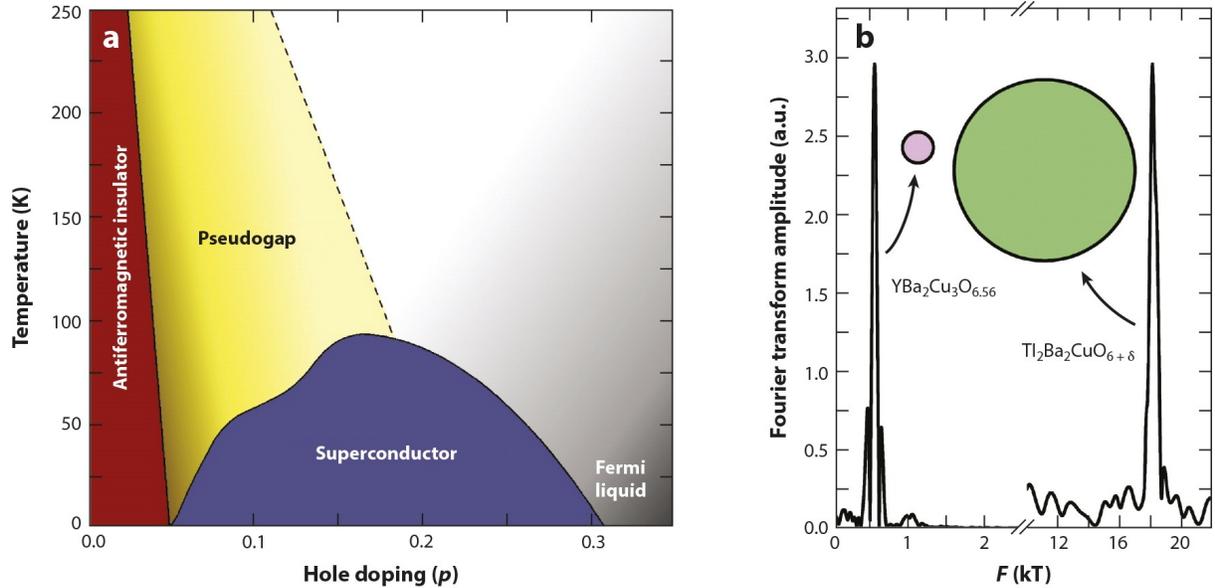

Figure 1: (a) Schematic phase diagram as a function of doping for hole-doped superconductors. The parent Mott insulating antiferromagnet at zero hole doping evolves to a $d$-wave superconducting ground state with increasing hole doping. Superconductivity forms a dome-like region, with the superconducting temperature increasing up to an optimal value in the vicinity of $p \approx 0.18$ hole doping and subsequently decreasing. The region of superconductivity below optimal doping is referred to as the underdoped regime, and the region of superconductivity above optimal doping is referred to as the overdoped regime. A Fermi liquid regime is observed beyond the end of the superconducting dome. The elevated temperature region above the superconducting dome in the overdoped regime is characterised by unusual properties, such as an absence of an antinodal density of states at the Fermi energy, and is known as the pseudogap regime. (b) Comparison between the small Fermi surface measured in the underdoped region;[26,35] data from[35] and the large Fermi surface measured in the overdoped regime;[28,61] data from.[28]

A particularly mysterious aspect of the high-temperature copper-oxide superconductors is the normal state out of which superconductivity evolves. The generic phase diagram of hole-doped cuprates shown in Figure 1a shows that high-temperature superconductivity is sandwiched between a Mott insulating parent compound and a Fermi liquid in the strongly overdoped regime. Although the $d$-wave nature of superconductivity in the copper-oxide superconductors is well characterised, the normal nonsuperconducting state on the underdoped side has remained a puzzle for over two decades.[3–9] In this underdoped regime, even when temperatures are elevated to suppress superconductivity and access the normal state, a mysterious gap in the density of states in the vicinity of the Fermi energy persists in the antinodal region of the Brillouin zone, known as the pseudogap.[10–12] Possible contributions to the pseudogap that have been theoretically proposed include pairing, charge, and magnetic, as well as other correlations of conventional or unconventional nature.[3–9,13–24] Experiments that probe the normal state using elevated temperatures to suppress superconductivity have revealed puzzling signatures of a state with seemingly little in common with a conventional Fermi liquid, leaving unresolved the origin of this state.

The discovery of quantum oscillations in the normal state of underdoped $YBa_2Cu_3O_{6+\delta}$[25] in 2007 transformed the landscape of the underdoped cuprates.[26] In this review, we examine progress made in the field of quantum oscillation measurements from their discovery to the insights they provide on the normal ground state of the underdoped cuprates. Quantum oscillations first reveal that the normal groundstate of the underdoped cuprates comprises primarily a small, quasi-two-dimensional, electron-like Fermi surface in contrast to the large hole-like Fermi surface calculated from band structure[27] and measured in the overdoped cuprates[28] (Figure 1b). Second, quantum oscillations reveal that the normal ground state in the underdoped cuprates has characteristics of a Fermi liquid. Third, careful study of the quantum oscillations reveal an electronic structure associated with a charge-ordered normal ground state in the underdoped regime of the cuprates. Fourth, the charge-ordered normal ground state extends over the majority of the underdoped regime in $YBa_2Cu_3O_{6+\delta}$ and appears to be characteristic of a translational symmetry broken groundstate that is universal to the hole-doped cuprates. Finally, the charge ordered normal groundstate is bounded on both the low-doped side and the high-doped side by two quantum critical points − indicated by a steep increase in effective mass − the locations of which underlie maxima of the superconducting two-subdome structure.

## 2   QUANTUM OSCILLATIONS IN CUPRATES

Quantum oscillations are a consequence of Landau quantisation of energy levels in an interacting electron system. In a semiclassical approximation, this leads to electrons executing cyclotron orbits confined to quantised Landau levels, the reciprocal space area of which is proportional to the magnetic field. As the applied magnetic field is increased, a discontinuous jump occurs in the density of states each time a Landau level exits the Fermi surface. All the physical properties of the material that are a function of the density of states therefore exhibit what are known as quantum oscillations i.e. oscillatory behaviour that is periodic in inverse magnetic fields. The frequency $F$ of quantum oscillations in inverse magnetic fields yields a measure of the Fermi surface area $A$ in momentum space, related through the Onsager relation: $F = \frac{\hbar}{2\pi e} A$. In addition, the temperature and the field dependencies of the amplitude of the oscillations yield the effective mass of the quasiparticle $m^*$ and the mean free path, respectively.[29–31]

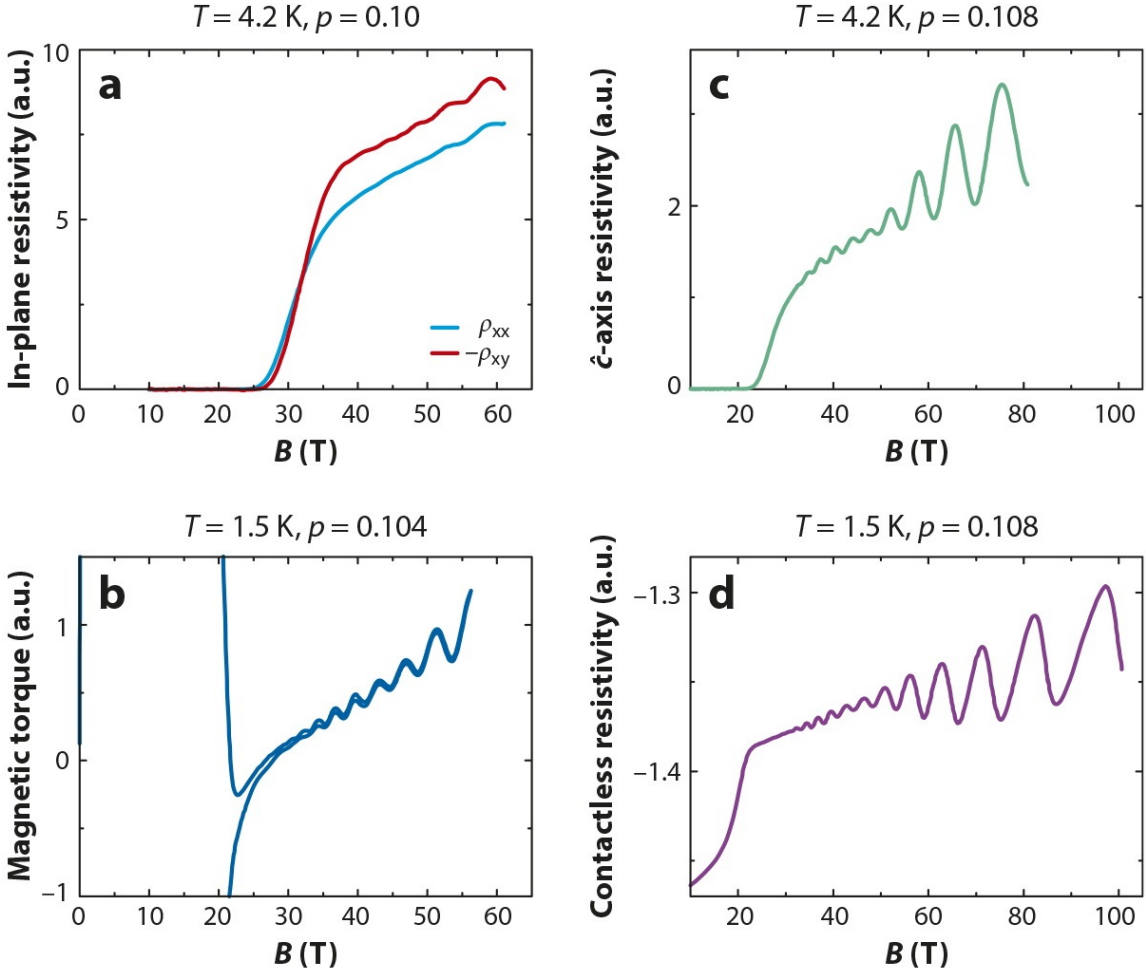

Figure 2: Quantum oscillations measured in underdoped $YBa_2Cu_3O_{6+\delta}$ by a variety of experimental techniques, including (a) in-plane four contact resistivity (data from[26]), (b) magnetic torque (data from[46]), (c) $\hat{c}$-axis four contact resistivity (data from[48]), and (d) contactless resistivity measured using a resonant proximity detection oscillator (data from[35]).

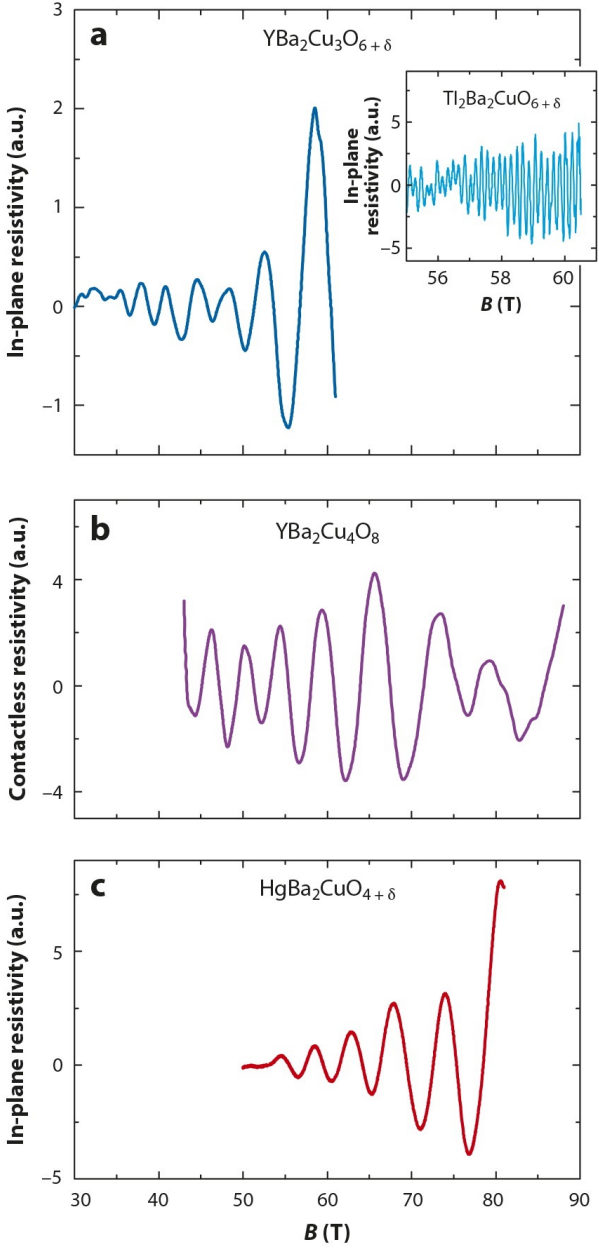

Figure 3: A comparison of quantum oscillations measured in three different families of underdoped cuprates shown in (a) $YBa_2Cu_3O_{6+\delta}$, (b) $YBa_2Cu_4O_8$ (data from[44]), and (c) $HgBa_2CuO_{4+\delta}$ (data from[45]). Quantum oscillations in overdoped $Tl_2Ba_2CuO_{6+\delta}$ (data from[28]) are shown in the inset.

## 2.1 Discovery of Quantum Oscillations

Quantum oscillations arise from the existence of well-defined quasiparticles obeying Fermi-Dirac statistics as in the Landau theory of a Fermi liquid. Therefore, it came as a great surprise when quantum oscillations were first discovered in the normal state of the underdoped cuprates, which was previously thought to have little relation to a Fermi liquid. Figure 2 shows quantum oscillation measurements in underdoped $YBa_2Cu_3O_{6+\delta}$ ortho II at a doping level $p \approx 0.11$ from different experimental probes: in-plane and Hall resistances,[26,32] magnetic torque,[33] $c$-axis resistance,[34] and contactless resistance measured by resonant oscillatory techniques.[35] Quantum oscillations in $YBa_2Cu_3O_{6+\delta}$ have also been observed in specific heat,[36] Nernst and Seebeck coefficients,[37] and thermal conductivity.[38] The common and main frequency of all probes $F = 530$ T correspond to an extremal area $A_F = 5.1$ nm$^{-2}$, which represents only 1.9 % of the first Brillouin zone. This is in sharp contrast with the high frequency of quantum oscillations found in overdoped $Tl_2Ba_2CuO_{6+\delta}$ at $p \approx 0.30$, assuming a similar phase diagram in the different families of hole-doped cuprates (see inset of Figure 3), where $F = 18,100$ T corresponds to a Fermi surface cross-section area, $A_F = 172.8$ nm$^{-2}$, which represents 65 % of the first Brillouin zone.[28]

Quantum oscillations have been observed in a wide range of doping of $YBa_2Cu_3O_{6+\delta}$, from $p \approx 0.09$ up to $p \approx 0.16$, over which the quantum oscillation frequency exhibits a subtle increase with doping.[39–41] Besides underdoped $YBa_2Cu_3O_{6+\delta}$, quantum oscillations have also been observed in the related compound $YBa_2Cu_4O_8$ ($T_c$ = 81 K corresponding to $p \approx 0.14$)[42–44] and more recently in another family of cuprates, namely $HgBa_2CuO_{4+\delta}$ ($T_c$ = 72 K corresponding to $p \approx 0.09$),[45] as shown in Figure 3. For the latter, the frequency of quantum oscillations $F$ = 840 T corresponds to a Fermi surface cross-section area $A_F = 8.0$ nm$^{-2}$, which represents 3 % of the first Brillouin zone.

## 2.2 Multiple Quantum Oscillation Frequencies

Understanding the origin of the small Fermi surface pocket revealed by quantum oscillations is challenging for multiple reasons. Firstly, it is unclear as to whether the traditional theory of quantum oscillations in quasi-two dimensional materials can be applied to these materials. Secondly, discerning the correct electronic structure associated with the observed quantum oscillations is challenging on the basis of calculated band structures.[27] Making the identification of the electronic structure more challenging is the experimental observation of multiple quantum oscillation frequencies, which could correspond either to different Fermi surface sections, to multiple extremal cross sections of the small Fermi surface, or to concentric split versions of a single Fermi surface. Figure 4 shows Fourier transforms of quantum oscillation measurements made using complementary experimental techniques of magnetic torque,[46] $c$-axis resistivity,[39] and contactless resistivity.[35] All of these techniques yield multiple peaks in the Fourier transform, with a central peak at 530 T bounded by two side Fourier peaks at 440 T and 620 T, the relative amplitude of which varies according to the experimental technique. Quantum oscillation frequencies originate generally from extremal cross-sectional areas of the Fermi surface but also from magnetic breakdown tunneling between extremal Fermi surface orbits. Alternative possibilities suggested for the origin of multiple frequencies observed in $YBa_2Cu_3O_{6+\delta}$ include fundamental neck and belly warping of the Fermi surface;[46–48] Fermi surface splitting due to effects such as bilayer coupling, accompanied by magnetic breakdown tunneling between split surfaces;[35] and adjoining Fermi surface pockets accompanied by magnetic breakdown tunneling between them.[49]

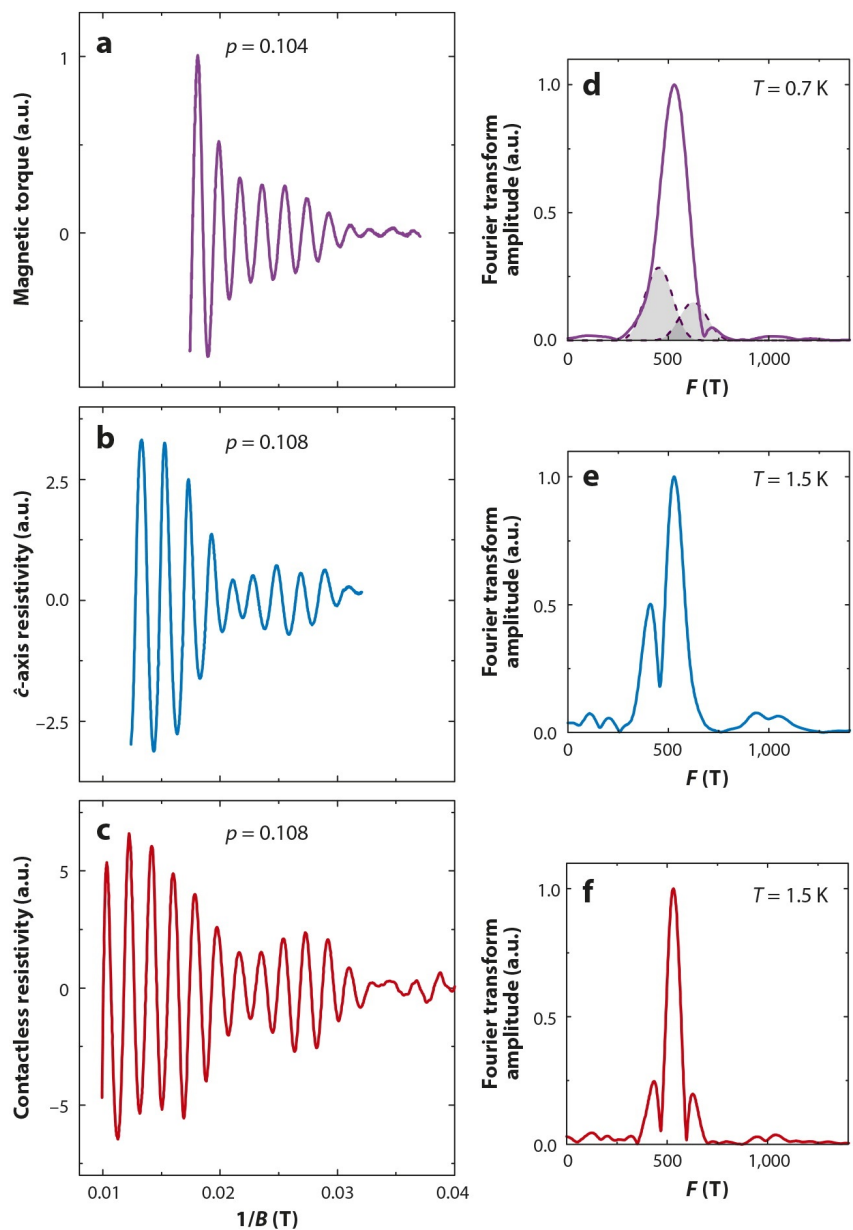

Figure 4: Measurement of multiple quantum oscillation frequencies in underdoped $YBa_2Cu_3O_{6+\delta}$ (left) and corresponding Fourier transform (right) using (a,d) magnetic torque (data from[46]), (b,e) $c$-axis resistivity (data from[39]), and (c,f) contactless resistivity (data from[35]). A dominant frequency of 530 T is observed, flanked by side frequencies and harmonics.

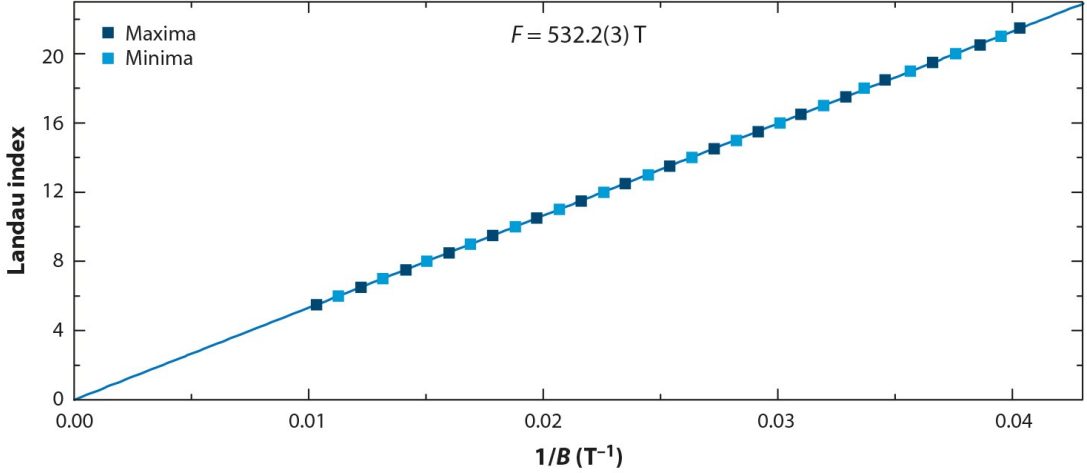

Figure 5: Maxima (dark blue) and minima (light blue) of quantum oscillations measured in $YBa_2Cu_3O_{6+\delta}$ (shown in Figure 4c) plotted as a function of inverse magnetic field. Excellent agreement is seen with linear behaviour. Quantum oscillations are therefore indicated to originate from Landau quantisation.[35,40]

## 3 FERMI LIQUID PROPERTIES

The previously widespread belief that the underdoped cuprates are non-Fermi liquids causes us to rigorously examine whether the quantum oscillations observed in underdoped $YBa_2Cu_3O_{6+\delta}$ exhibit characteristics, such as Landau quantisation and Fermi-Dirac statistics, of conventional metals. Alternative explanations to quantum oscillations from a Landau-quantised Fermi liquid that have been put forward include, for example, quantum oscillations from the vortex lattice, which would yield a different magnetic field dependence.[50–52]

## 3.1 Landau Quantisation

In order to verify Landau quantisation, quantum oscillations are examined over a broad range in magnetic field.[35,40] Figure 2d shows quantum oscillations measured from 22 T to 100 T. Inverse magnetic fields at which quantum oscillation maxima occur are labelled by integers, and inverse magnetic fields at which quantum oscillation minima occur are labelled by half-integers. On plotting the integer and half-integer Landau indices as a function of inverse magnetic field (Figure 5), robust linear behaviour is seen, verifying Landau quantisation.[29–31]

## 3.2 Fermi-Dirac Statistics

Quantum oscillations are an excellent tool to detect the statistical distribution of the underlying particles. The temperature dependence of the quantum oscillation amplitude provides a direct measure of the underlying particle statistics, which has a characteristic temperature dependence. For instance, the distinctive step in energy separating occupied from unoccupied states at $T = 0$ in the case of Fermi-Dirac statistics is smeared at higher temperatures (Figure 6a). Consequently, particles with Fermi-Dirac statistics are expected to yield a quantum oscillation amplitude that decreases with increasing temperature, which is known as the Lifshitz Kosevich form.[29–31] The line in Figure 6c shows the characteristic temperature dependence of the derivative of the Fermi-Dirac statistical distribution.[53] Figure 6b shows quantum oscillations measured at finely spaced temperatures over a temperature range between 1.1 K and 16 K. The symbols in 6c show the inverse Fourier transform of the measured temperature dependence of the quantum oscillation amplitude between 0.1 K and 18 K. The excellent agreement in Figure 6c indicates the particle statistics associated with the observed quantum oscillations in underdoped $YBa_2Cu_3O_{6+\delta}$ as being Fermi-Dirac in character.[53] Strikingly, therefore, quantum oscillation measurements reveal a normal Fermi liquid ground state in the underdoped regime of the cuprates characterised by Landau quantisation and Fermi-Dirac statistics.

# 4 FERMI SURFACE RECONSTRUCTION

## 4.1 Experimental Evidence

Given Fermi liquid behaviour exhibited by quantum oscillations in underdoped $YBa_2Cu_3O_{6+\delta}$, it is reasonable to interpret these using the traditional framework used for metallic quasi-two dimensional systems.[29–31] Within such a framework, the first measurements of quantum oscillations in underdoped $YBa_2Cu_3O_{6+\delta}$ reveal two striking pieces of information.[26] Firstly, the low frequency of measured quantum oscillations is equivalent to a Fermi surface area of only $\approx 2\%$ of the Brillouin zone. Secondly, the sign of the Hall resistivity in which the quantum oscillations occur is found to be negative. Indeed, Figure 7b shows the temperature dependence of the normal state Hall coefficient measured at high field in underdoped $YBa_2Cu_3O_{6+\delta}$. As demonstrated in References,[54,55] the negative sign of the Hall coefficient at low temperature is shown to be a property of the normal state. The most natural explanation for the negative Hall coefficient is the presence of an electron pocket in the Fermi surface. This is consistent with measurements of the Seebeck coefficient (or thermopower), the sign of which is controlled by the carrier type (positive for holes, negative for electrons). As seen in Figure 7c, the Seebeck coefficient measured at high field in underdoped $YBa_2Cu_3O_{6+\delta}$ undergoes a change of sign, from positive at high temperature to negative at low temperature, similar to the sign change reported in the Hall coefficient $R_H$, indicating an electron-like surface in the electronic structure of the normal underdoped cuprates.[37,56]

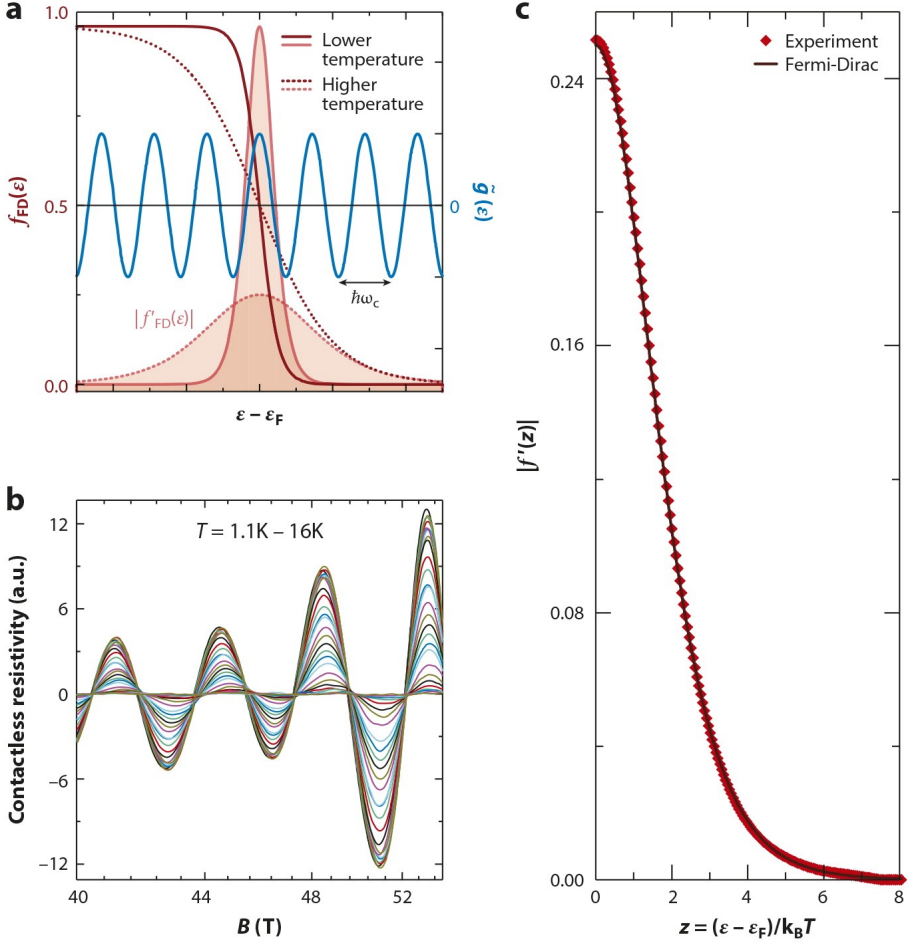

Figure 6: (a) Schematic of the Fermi-Dirac distribution $f_{\text{FD}} = (1 + e^z)^{-1}$ [where $z = (\varepsilon - \varepsilon_{\text{F}})/k_{\text{B}}T$]. The $T$-dependent step in occupation number (lines, dotted for higher temperatures) causes the oscillatory density of states $\tilde{g} = g_0 e^{i2\pi\varepsilon/\hbar\omega_{\text{c}}}$ (assuming that $g_0$ is approximately constant on the scale of the cyclotron energy $\hbar\omega_{\text{c}} = \hbar eB/m^*$) shown by the sinusoidal line (blue) to be thermally smeared by the derivative of the probability distribution $|f'_{\text{FD}}(z)| = 1/2(1 + \cosh z)$ (shaded regions). The consequent reduction in amplitude is equivalent to a Fourier transform of $|f'_{\text{FD}}(z)|$, yielding oscillations $\propto e^{i(2\pi F/B)}$ periodic in $1/B$ modulated by a $T$-dependent prefactor $a(T) = a_0 \pi\eta/\sinh\pi\eta$ (where $\eta = 2\pi k_{\text{B}}Tm^*/eB$ and $a_0$ is a constant). The quantum oscillatory magnetisation and resistivity can be expressed in terms of the above thermally averaged density of states, hence the same thermal amplitude factor $a(T)$. (b) Magnetic quantum oscillations (after background polynomial subtraction) measured in YBa$_2$Cu$_3$O$_{6.56}$ ($p \approx 0.108$). This restricted interval in $B = |\mathbf{B}|$ furnishes a dynamic range of $\approx 50$ dB over the range of measured temperatures of 1.1, 1.2, 1.4, 1.7, 1.8, 2.0, 2.1, 2.2, 2.6, 3.0, 3.5, 3.9, 4.0, 4.5, 5.0, 5.5, 6.0, 6.5, 7.0, 8.0, 9.0, 10.0, 11.0, 12.0, 14.0, 15.0, 16.0 K. (c) Inverse Fourier transform (red diamonds) of the amplitude of the oscillations versus $z$. Its comparison with the Fermi-Dirac distribution (black line) shows excellent agreement. Adapted from Reference[53]

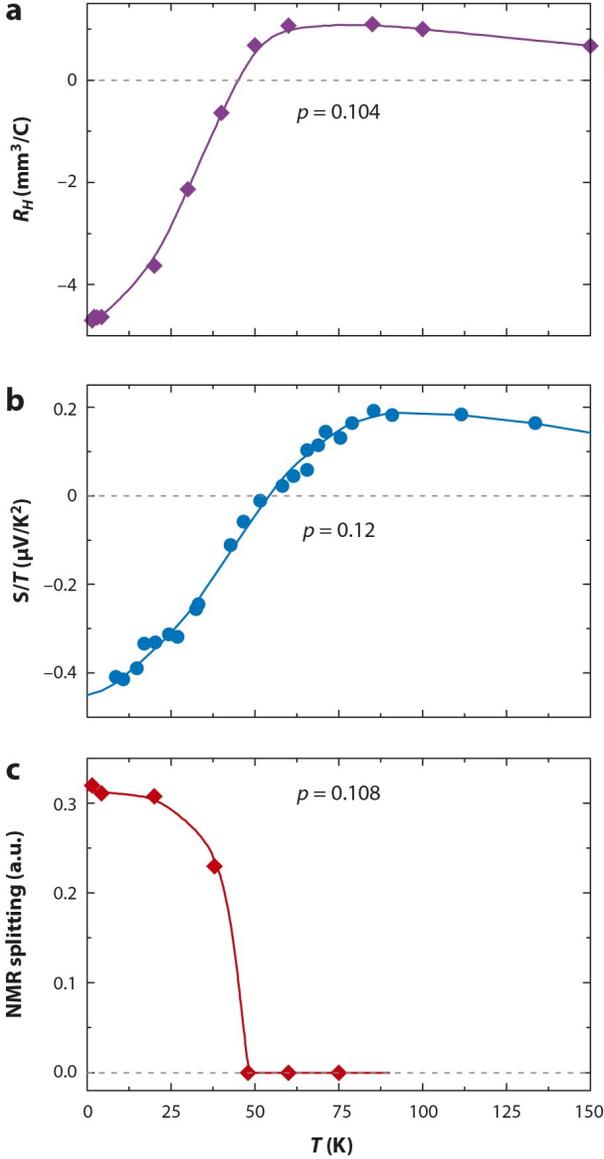

Figure 7: Temperature dependence of (a) the Hall effect (adapted from Reference[54]) and (b) the Seebeck coefficient (adapted from Reference[56]) in $YBa_2Cu_3O_{6+\delta}$ measured at high fields. (c) Charge order identified by the onset of line splitting of the high-frequency Cu2F satellite in nuclear magnetic resonance measurements at a magnetic field of 28.5 T.[69] The temperature below which line-splitting (shown in (a)) is resolved is similar to the temperature below which the Hall and Seebeck effect become negative,[54,56] indicating Fermi surface reconstruction (shown in (b) and (c)).

## 4.2 Alternative Fermi surface Models

Two broad sets of alternative proposals have been put forward to explain the origin of the drastic reduction in Fermi surface size and its transformation to electron-like character. In the first set of proposals, the small Fermi pocket arises from a region of band structure distinct from the superconducting $CuO_2$.[27] An example of this category of proposals is one in which a small Fermi pocket arises from hybridisation of the CuO chain and BaO bands.[57–59] This scenario, however, faces multiple difficulties: (a) It is challenging for such a scenario to explain the electron-like character of the small Fermi surface pocket[54,60] and (b) quantum oscillations with similar frequency have been detected in underdoped $HgBa_2CuO_{4+\delta}$, a compound free of CuO chains.[45] In the second set of proposals, which are more relevant in explaining the observed quantum oscillations, the small Fermi pocket arises from a form of translational symmetry breaking order that reconstructs the large hole-like Fermi surface predicted from band structure and observed in the overdoped cuprates.[27,28,61] Examples of models proposed in this category[62–68] are illustrated in Figure 8. They include translational symmetry breaking from $d$-density order[16] or commensurate antiferromagnetic order (Figure 8a); stripe order, e.g., a combination of charge and spin order[62] (Figure 8b); charge order in conjunction with nematic order[64] (Figure 8c); and others. Characteristic of all these models is the occurrence of electron pockets at the antinodal location, potentially accompanied by hole pockets at the nodal location. Here, nodal refers to the region of momentum space in which the $d$-wave superconducting gap is minimised, and antinodal refers to the region of momentum space in which the $d$-wave superconducting gap is maximised. Such a positioning of the Fermi pocket in the antinodal region of the Brillouin zone proved challenging to understand given the large gap in the antinodal density of states at the Fermi energy[4,10] measured by complementary experiments, such as photoemission,[11] optical conductivity,[12] and others.

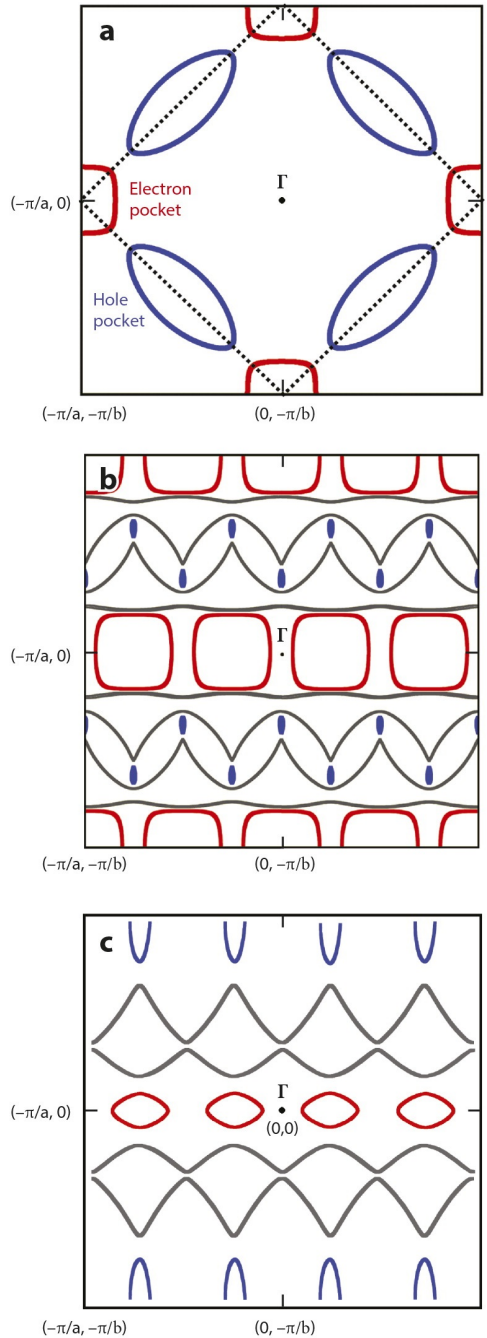

Figure 8: Alternative Fermi surface proposals in which hole pockets (blue) and electron pockets (red) arise at the nodal and antinodal locations of the Brillouin zone, respectively. (a) Fermi surface reconstruction by antiferromagnetic or $d$-density wave order.[63] (b) Fermi surface reconstruction by charge-spin stripes.[62] (c) Fermi surface reconstruction by unidirectional charge accompanied by nematic order. Figure adapted from Reference.[64]

# 5 EVIDENCE FOR TRANSLATIONAL SYMMETRY BREAKING

## 5.1 Observation of Charge Order by Nuclear Magnetic Resonance

After the first two decades in which no signatures of translational symmetry breaking were observed in underdoped $YBa_2Cu_3O_{6+\delta}$, the observation of a small Fermi surface by quantum oscillation measurements motivated a renewed search using a variety of complementary techniques. High fields nuclear magnetic resonance (NMR) measurements[69] of copper nuclei performed under conditions for which quantum oscillations are observed revealed a line-splitting for the planar sites which lie below oxygen-filled chains (and the absence thereof for sites below empty chains). This splitting involves a differentiation of the quadrupole frequency (due to the interaction of the nucleus with electric field gradients) of the planar sites, therefore providing direct evidence for charge order or quasi-order in the $CuO_2$ planes. As seen in Figure 7c, the charge differentiation is observed by NMR below a temperature of the order of $50 \pm 10$ K for underdoped $YBa_2Cu_3O_{6+\delta}$ ($p \approx 0.11$); in this region, the Hall constant $R_H$ becomes negative. This effect is observed above a threshold magnetic field, suggesting an interplay of charge order with superconductivity at low temperatures.[70]

Notably, spin order was not observed by these NMR measurements in the temperature and magnetic field regime explored. This is consistent with multiple spin zeros from Zeeman splitting observed in fundamental quantum oscillations measured over a range of tilt angles between the magnetic field and the crystalline $\hat{c}$-axis.[40,48]

## 5.2 Charge-Ordering Wavevectors

Complementary resonant soft X-ray scattering,[71,72] hard X-ray scattering,[73] and inelastic X-ray scattering[74] experiments performed in underdoped $YBa_2Cu_3O_{6+\delta}$ revealed charge order (or quasi-order) at zero magnetic field. The charge order is found to be characterised by orthogonal and incommensurate wavevectors $\mathbf{Q_x} = 2\pi(\pm\frac{\delta_1}{a}, 0, \pm\frac{1}{2c})$ and $\mathbf{Q_y} = 2\pi(\pm\frac{\delta_2}{a}, 0, \pm\frac{1}{2c})$ with $\delta_1 \approx \delta_2 \approx 0.3$. The intensity of the charge modulation increases from temperatures of the order of 150 K down to $T_c$ and then decreases in the superconducting state. Below $T_c$, the application of a magnetic field weakens superconductivity and enhances the charge order. This temperature dependent behaviour may be interpreted as the suppression of superconductivity by a magnetic field, which enhances charge correlations. Additionally, high-field sound velocity measurements in underdoped $YBa_2Cu_3O_{6+\delta}$ show a thermodynamic signature of the charge-ordering phase transition which indicate biaxial charge modulation based on a group theory analysis.[75] Further signatures of charge order are revealed in optical reflectometry experiments in $YBa_2Cu_3O_{6+\delta}$,[76] and $La_{2-x}Sr_xCuO_4$.[77] Other than the La-based cuprates in which stripe order was previously identified,[78] charge order has now been detected in many cuprate families, namely in $Bi_2Sr_2CaCu_2O_{8+\delta}$,[79] $Bi_2Sr_{2-x}La_xCuO_{6+\delta}$,[80] $HgBa_2CuO_{4+\delta}$,[81] and $La_{2-x}Sr_xCuO_4$.[82,83] Therefore, indications are, that charge order (or quasi-order) is a univer-sal feature of the phase diagram of the copper-oxide superconductors. We note that for Fermi surface reconstruction, a superlattice need not be strictly long-range or static, but it must not be fluctuating over a range much smaller than the cyclotron radius, nor with a frequency much larger than the cyclotron frequency.

# 6 NODAL FERMI SURFACE FROM CHARGE ORDER

## 6.1 Indication of the location of the electron pocket

A careful analysis of quantum oscillation measurements over a broad range of magnetic field and angle has provided important information about the location of the pocket in momentum space. Figure 9 shows quantum oscillations measured in underdoped YBa$_2$Cu$_3$O$_{6+\delta}$ up to magnetic fields of 85 T at tilt angles to the magnetic field up to 71°.[84] Although the beat pattern at a 0° tilt angle could arise either from a fundamental neck and belly warping[46–48] or a Fermi surface splitting,[35] the evolution in beat pattern with tilt angle enables us to distinguish between these possibilities. In particular, the absence of a Yamaji resonance[30,85] in amplitude at high tilt angles is inconsistent with a sizeable neck and belly warping. Instead, the angular dependence of measured quantum oscillations points to quasi-two dimensional Fermi surface cylinders that are split by a finite bilayer or spin-orbit coupling, in which the fundamental neck and belly geometry is replaced by a staggered twofold warping. This unique form of Fermi surface warping is unsupported by the original primitive orthorhombic Brillouin zone in underdoped YBa$_2$Cu$_3$O$_{6+\delta}$. Just this form of warping, would, however, arise upon transformation of the original Brillouin zone into a body-centred orthorhombic Brillouin zone by a superlattice with ordering wavevectors $\mathbf{Q_x} = 2\pi(\pm\frac{\delta}{a}, 0, \pm\frac{1}{2c})$ and $\mathbf{Q_y} = 2\pi(0, \pm\frac{\delta}{a}, \pm\frac{1}{2c})$. Under such a Fermi surface reconstruction, a small Fermi surface with staggered twofold warping would arise in the vicinity of the nodal planes of the Brillouin zone see Figure 10.[40,86,87] Given complementary findings of ordering wavevectors $\mathbf{Q_x}$ and $\mathbf{Q_y}$ associated with charge order,[71–74] we are able to conclude that the underdoped normal ground state comprises primarily a nodal electron-like Fermi pocket created by charge order.[84]

## 6.2 Fermi Surface Model

Quantum oscillation and transport measurements in conjunction with measurements of translational symmetry breaking observed by complementary spectroscopies converge toward a Fermi surface reconstruction by charge order. A model of charge order with finite amplitude components of wavevectors $\mathbf{Q_x}$ and $\mathbf{Q_y}$ has been proposed in References[40,86–88] that can, together with other effects discussed below, suppress the Fermi surface at the antinodal region and create electron-like pockets from the nodal density of states at the Fermi energy in place of the large hole-like Fermi surface characterising the overdoped state.[28] This model of nodal electron pockets for the underdoped state is consistent not only with the Fermi surface geometry inferred from quantum oscillations but also the low quasiparticle density of states at the Fermi level inferred from the measured low-temperature specific heat coefficient[36] and the observation of chemical potential oscillations,[89] considerations of which indicate that if other pockets were to exist, such as the hole pockets shown in Figure 10b.[87] The observation of a negative Hall coefficient[26,54] and the subtle increase in Fermi pocket area with increasing hole doping[39,94] is further explained by this model.[87] An open question, however, surrounds the underlying mechanism responsible for charge order characterised by ordering wavevectors $\mathbf{Q_x} = 2\pi(\pm\frac{\delta}{a}, 0, \pm\frac{1}{2c})$ and $\mathbf{Q_y} = 2\pi(0, \pm\frac{\delta}{a}, \pm\frac{1}{2c})$. Various theoretical possibilities have been suggested, including those involving interplay between magnetism, superconductivity, and charge order.[90,91] It is also of interest to consider whether the nature of charge order could be different in zero and high magnetic fields.[92]

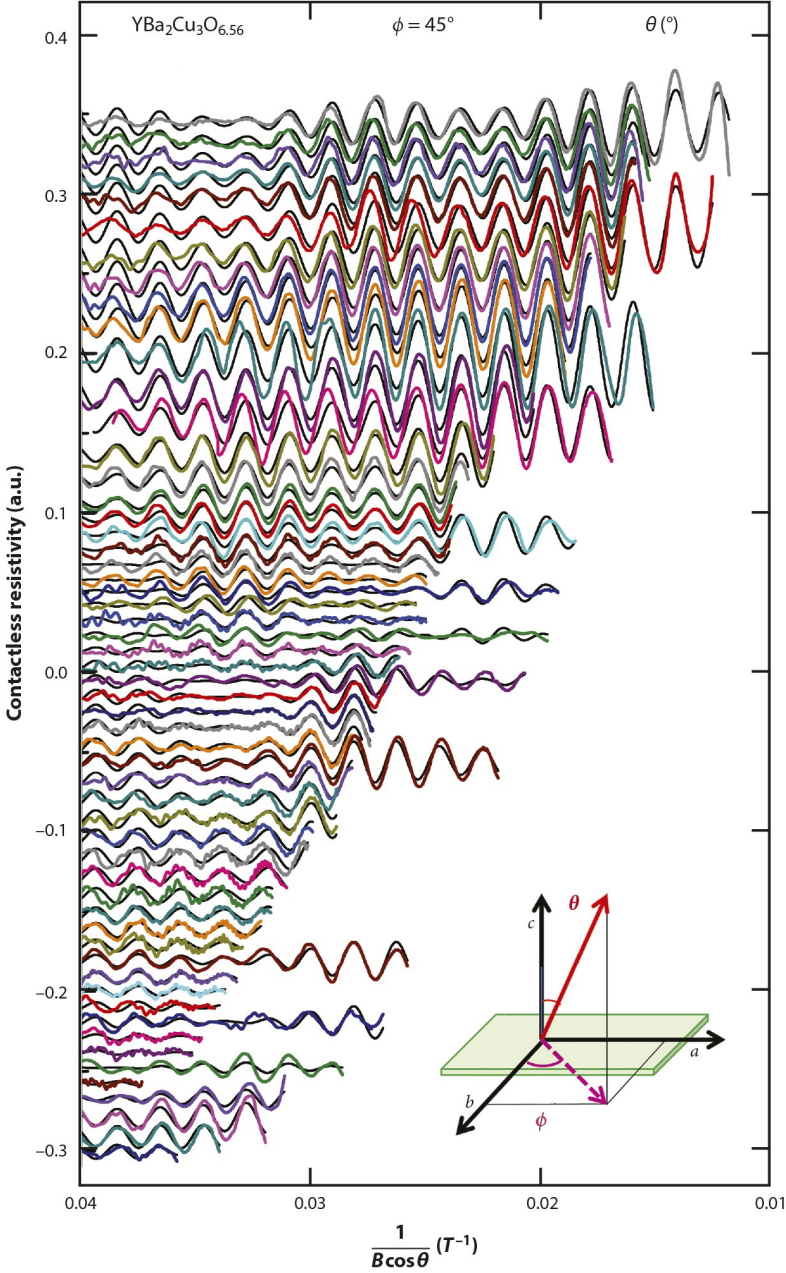

Figure 9: Quantum oscillations measured in YBa$_2$Cu$_3$O$_{6.56}$ ($p \approx 0.108$) at various angles of inclination of the magnetic field: $\theta = 0$, 1.3, 11.3, 12, 16.3, 18, 21.3, 26.3, 31.3, 36.3, 38, 41.3, 45.2, 46.3, 48, 49, 49.4, 50.1, 50.6, 51.4, 51.5, 52, 52.3, 52.5, 52.9, 53.1, 54.4, 54.9, 55.5, 56, 56.2, -56.95, 57.2, -57.4, -58.15, 58.2, -59.4, 59.6, 60.6, 61.2, -61.4, 61.7, 62.5, 62.6, -62.7, -63.2, 63.4, 63.7, -64.1, 64.5, 65.5, 66, 66.3, 68.1, 69.4 and 70.6°. In black are simulations of a staggered twofold split Fermi surface model with a splitting of 90 T and a twofold warping of 14 T, which shows good agreement with the experimental data (coloured). Adapted from Reference.[84]

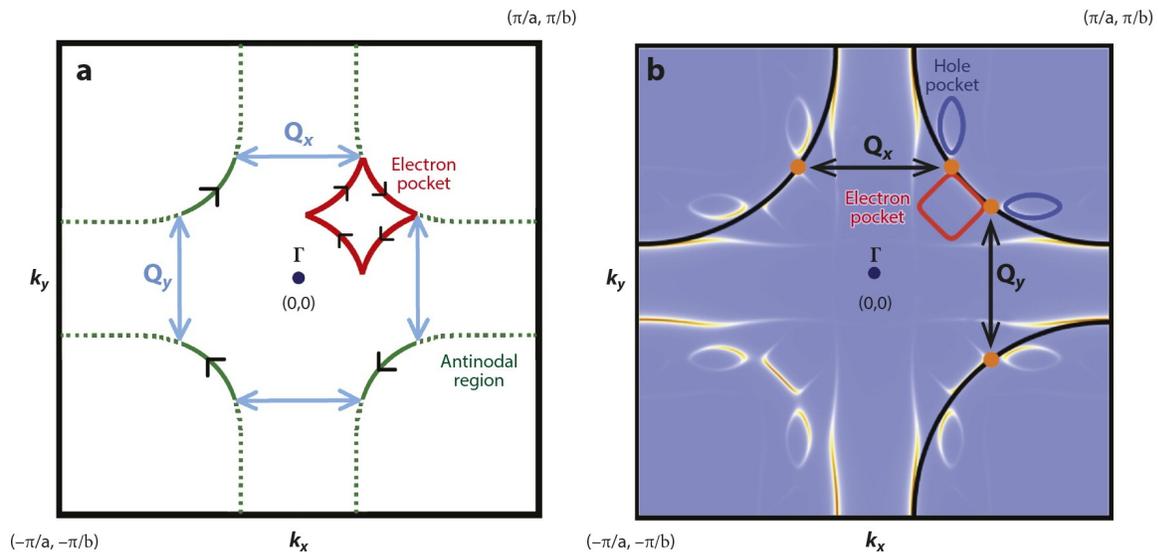

Figure 10: Charge order with wavevectors $\mathbf{Q}_x$ and $\mathbf{Q}_y$ of similar or dissimilar amplitudes that yields a nodal electron pocket (red). A negative Hall effect is expected from such a pocket at high magnetic fields. The dotted green lines represent antinodal regions, which are suppressed due likely to a combination of charge correlations in conjunction with effects such as pairing or magnetic correlations.[40,86] Figure adapted from References.[40,86] (b) Quantum mechanical calculation of an electron spectral function. Charge order with $d$-wave symmetry (i.e., bond order) with wavevectors $\mathbf{Q}_x$ and $\mathbf{Q}_y$ gives similar results to those found in panel a, yielding a nodal electron pocket (red) accompanied by small adjoining hole pockets (blue).[87] Figure adapted from Reference.[87]

## 6.3 Fermi Surface Evolution with Hole Doping

Measurement of quantities such as the superconducting temperature[93] and upper critical magnetic field[38,94] indicates that the superconducting regime comprises a two-subdome superconducting structure (Figure 11a-c). Quantities such as the step in heat capacity at the superconducting transition,[95] and others,[96] indicate a putative quantum critical point underlying each of the subdome maxima (Figure 11), similar to other materials families such as the heavy fermion $CeCu_2Si_2$.[98]

An interesting question relates to the connection between the Fermi surface and the two-dome superconducting structure. Quantum oscillations have been observed above a lower hole doping of $p \approx 0.09$, and up to a hole doping of $p \approx 0.16$ close to optimal doping in $YBa_2Cu_3O_{6+\delta}$.[39,41,94,99] The effective quasiparticle mass of the observed small Fermi surface is seen to increase steeply toward both the underdoped limit and the optimal doping limit. On extrapolating the inverse effective mass as a function of hole doping, a collapse in inverse effective mass (i.e. an effective mass divergence) is indicated in the vicinity of both a lower quantum critical point at doping $p_1 \approx 0.085$[99] underlying the maximum of the lower subdome and a higher quantum critical point at doping $p_2 \approx 0.18$[41,94] underlying the maximum of the higher subdome ( Figure 11d). Charge order as observed by X-ray diffraction spans the hole-doping regime between these two quantum critical points, over which Fermi surface reconstruction is observed by quantum oscillations.[100,101] Magnetic order is observed to onset at dopings below the lower critical point,[102–104] and charge order as observed by X-ray diffraction vanishes for dopings above the higher quantum critical point.[100]

## 7 BROADER IMPLICATIONS

From the array of quantum oscillation measurements performed to date in the underdoped cuprates, we are able to conclude that the normal ground state throughout the majority of the underdoped regime in $YBa_2Cu_3O_{6+\delta}$ is characterised by a nodal electron Fermi pocket created by Fermi surface reconstruction by charge order. This normal ground state is likely universal in a majority of hole-doped cuprates.

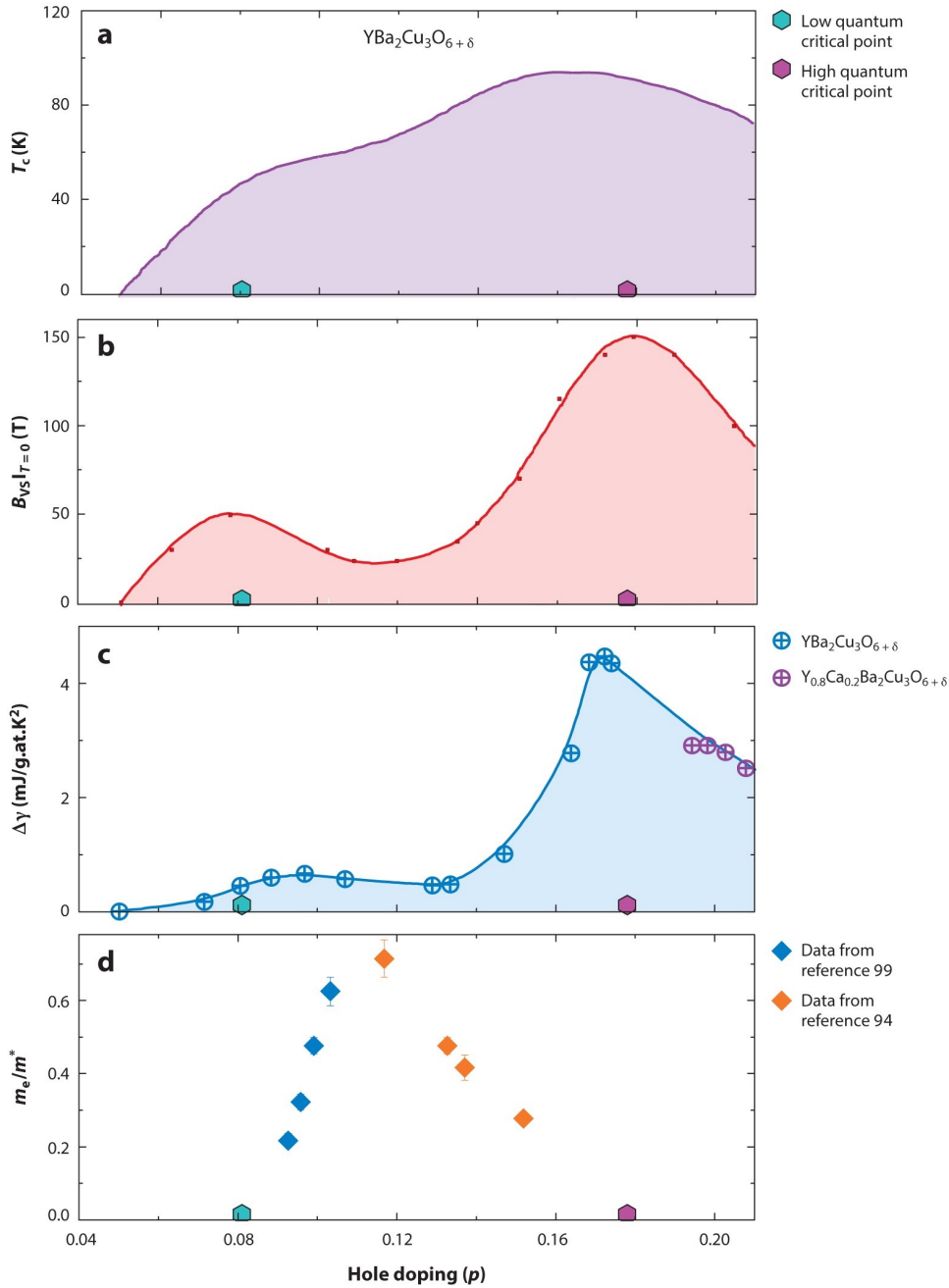

Figure 11: Two subdome structure of (a) the superconducting temperature[93] and (b) the upper critical magnetic field (data from[38]) in $YBa_2Cu_3O_{6+\delta}$. (c) Enhancement of the step height in specific heat at the superconducting transition underlying each of the subdome maxima (data from [84] and [95]). A lower and higher quantum critical point are indicated beneath the each of the maxima of the two superconducting subdomes. (d) A rapid enhancement in effective mass measured by quantum oscillations is seen to extrapolate to a mass divergence at the putative quantum critical point underlying the maxima of each of the lower and higher superconducting subdomes (data from[94,99]).

An important outstanding question pertains to the origin of the antinodal gap that characterises the pseudogap state.[10–12] Quantum oscillation measurements that reveal a nodal electron Fermi pocket leave open two scenarios. In the more likely scenario, the antinodal region is gapped by pairing and/or magnetic correlations prior to the onset of charge order (or quasi-order).[105–110] In an alternative scenario, the antinodal region is gapped by strong charge correlations.[73,111]

While some open questions remain, there is no doubt that the resolution of the electronic structure in the normal ground state of the underdoped cuprates by quantum oscillations, has, to a large extent demystified this regime, positioning us much closer to finally solving the puzzle of high temperature superconductivity in the copper-oxide materials.

# 8 ACKNOWLEDGEMENTS

We wish to warmly thank our collaborators, including A. Audouard, S. Badoux, F. F. Balakirev, N. Barisic, D. A. Bonn, A. Carrington, N. Doiron-Leyraud, P. A. Goddard, M. Greven, W. N. Hardy, N. Harrison, N. Hussey, C. Jaudet, M. -H. Julien, F. Laliberté, D. Leboeuf, J. Levallois, R. Liang, G. G. Lonzarich, C. H. Mielke, M. Nardone, B. J. Ramshaw, W. Tabis, L. Taillefer, B. Vignolle, and D. Vignolles. C.P. acknowledges research support from the CNRS, the ANR project SUPERFIELD, the project highTc of the labex NEXT and EMFL. S.E.S. acknowledges support from the Royal Society, King?s College Cambridge, the Winton Programme for the Physics of Sustainability, and the European Research Council grant number FP/2007-2013/ ERC Grant Agreement number 337425.

# References and Notes

[1] Bardeen J, Cooper LN, Schrieffer JR. 1957. *Phys. Rev* 108:1175–1204

[2] Bednorz JG, Muller KA. 1986. *Z. Phys. B* 64:189–193

[3] Chubukov AV, Morr DK. 1997. *Phys. Rep.* 288:355–387

[4] Norman M, Pépin C. 2003. *Rep. Prog. Phys.* 66:1547-1610

[5] Kivelson SA, Bindloss IP, Fradkin E, Oganesyan V, Tranquada JM, Kapitulnik A, Howald C. 2003. *Rev. Mod. Phys.* 75:1201–1241

[6] Sachdev S. 2003. *Rev. Mod. Phys.* 75:913–932

[7] Lee PA, Nagaosa N, Wen X-G. 2006. *Rev. Mod. Phys.* 78:17–85

[8] Yang K-Y, Rice TM, Zhang F-C. 2006. *Phys. Rev. B* 73:174501

[9] Anderson PW, Lee PA, Randeria M, Rice TM, Trivedi N, Zhang FC. 2006. *J. Phys.: Condens. Matter* 16:R755-R769

[10] Timusk T, Statt BW. 1999. *Rep. Prog. Phys.* 62:61–122

[11] Norman MR, Ding H, Randeria M, Campuzano JC, Yokoya T, Takeuchi T, Takahashi T, Mochiku T, Kadowaki K, Guptasarma P, Hinks DG. 1998. *Nature* 392:157–160

[12] Basov DN, Timusk, T. 2005. *Rev. Mod. Phys.* 77:721–779

[13] Anderson PW. 1987. *Science* 235:1196–1198

[14] Nagaosa N, Lee PA. 1992. *Phys. Rev. B* 45:966–970

[15] Emery V, Kivelson SA. 1995. *Nature* 374:434–437

[16] Chakravarty S, Laughlin RB, Morr DK, Nayak C. 2001. *Phys. Rev. B* 63:094503

[17] Wang Z-Q, Kotliar G, Wang X-F. 1990. *Phys. Rev. B* 42:8690–8693

[18] Varma CM. 1999. *Phys. Rev. Lett.* 83:3538–3542

[19] Castellani C, Di Castro C, Grilli M. 1995. *Phys. Rev. Lett.* 75:4650–4654

[20] Abanov A, Chubukov AV, Schmalian J. 2003. *Adv. Phys.* 52:119–218

[21] Vojta M. 2009. *Adv. Phys.* 58:699–820

[22] Zaanen J, Chakravarty S, Senthil T, Anderson P, Lee PA, Schmalian JA, Imada M, Pines D, Randeria M, Varma C, Vojta M, Rice TM. 2006. *Nat. Phys.* 2:138–143

[23] Scalapino DJ. 2012. *Rev. Mod. Phys.* 84:1383–1417

[24] Laughlin RB. 2014. *Phys. Rev. B* 89:035134

[25] Liang R, Bonn DA, Hardy WN. 2012. *Philosophical Magazine* 92:2563–2581

[26] Doiron-Leyraud N, Proust C, LeBoeuf D, Levallois J, Bonnemaison JB, Liang R, Bonn DA, Hardy WN, Taillefer L. 2007. *Nature* 447:565-568

[27] Andersen OK, Liechtenstein AI, Jepsen O, Paulsen F. 1995. *J. Phys. Chem. Solids* 56:1573–1591

[28] Vignolle B, Carrington A, Cooper RA, French MMJ, Mackenzie AP, Jaudet C, Vignolles D, Proust C and Hussey NE. 2008. *Nature* 455:952-955

[29] Shoenberg D. 1984. *Magnetic Oscillations in Metals* (Cambridge University Press, Cambridge)

[30] Bergemann C, Mackenzie AP, Julian SR, Forsythe D, Ohmichi E. 2003. *Adv. Phys.* 52:639–725

[31] Singleton J, Shoenberg D. 2001. *Band Theory and Electronic Properties of Solids*, (Oxford University Press).

[32] Jaudet C, Levallois J, Audouard A, Vignolles D, Vignolle B, Liang R, Bonn DA, Hardy WN, Hussey NE, Taillefer L, and Proust C. 2009. *Physica B* 404:354–358

[33] Jaudet C, Vignolles D, Audouard A, Levallois J, LeBoeuf D, Doiron-Leyraud N, Vignolle B, Nardone M, Zitouni A, Liang R, Bonn DA, Hardy WN, Taillefer L, Proust C. 2008. *Phys. Rev. Lett.* 100:187005

[34] Vignolle B, Ramshaw BJ, Day J, LeBoeuf D, Lepault S, Liang R, Hardy WN, Bonn DA, Taillefer L, Proust, C. 2012. *Phys. Rev. B* 85:224524

[35] Sebastian SE, Harrison N, Liang R, Bonn DA, Hardy WN, Mielke C, Lonzarich GG. 2012. *Phys. Rev. Lett.* 108:196403

[36] Riggs SC, Vafek O, Kemper JB, Betts JB, Migliori A, Balakirev FF, Hardy WN, Liang R, Bonn DA, Boebinger GS. 2011. *Nature Phys.* 7:332-335

[37] Laliberté F, Chang J, Doiron-Leyraud N, Hassinger E, Daou R, Rondeau M, Ramshaw BJ, Liang R, Bonn DA, Hardy WN, Pyon S, Takayama T, Takagi H, Sheikin I, Malone L, Proust C, Behnia K, Taillefer L. 2011. *Nature Commun.* 2:432

[38] Grissonnanche G *et al*. 2014. *Nature Commun.* 5:3280

[39] Vignolle B, Vignolles D, Julien M-H, Proust C. 2013. *Comptes Rendus Physique* 14:39–52

[40] Sebastian SE, Harrison N, Lonzarich GG. 2012. *Rep. Prog. Phys.* 75:102501

[41] Singleton J, de la Cruz C, McDonald RD, Li S, Altarawneh M, Goddard PA, Franke I, Rickel D, Mielke CH, Yao X, Dai P. 2010. *Phys Rev Lett.* 104:086403

[42] Yelland EA, Singleton J, Mielke CH, Harrison N, Balakirev FF, Dabrowski B, Cooper JR. 2008. *Phys. Rev. Lett.* 100:047003

[43] Bangura AF, Fletcher JD, Carrington A, Levallois J, Nardone M, Vignolle B, Heard PJ, Doiron-Leyraud N, LeBoeuf D, Taillefer L, Adachi S, Proust C, Hussey NE. 2008. *Phys. Rev. Lett.* 100:047004

[44] Tan BS, Zhu Z, Harrison N, Srivastava A, Ramshaw BJ, Sabok SA, Dabrowski B, Sebastian SE. 2014. *Proc. Natl Acad. Sci. USA* in press

[45] Barišić N, Badoux S, Chan MK, Dorow C, Tabis W, Vignolle B, Yu G, Béard J, Zhao X, Proust C, Greven M. 2013. *Nature Physics* 9:761–764

[46] Audouard A, Jaudet C, Vignolles D, Liang R, Bonn DA, Hardy WN, Taillefer L, Proust C. 2009. *Phys. Rev. Lett.* 103:157003

[47] Sebastian SE, Harrison N, Altarawneh MM, Goddard PA, Mielke CH, Liang R, Bonn DA, Hardy WN, Lonzarich GG. 2010. *Phys.Rev. B* 81:214524

[48] Ramshaw BJ, Vignolle B, Day J, Liang R, Hardy WN, Proust C, Bonn DA. 2011. *Nature Phys.* 7:234-238

[49] Doiron-Leyraud N, Badoux S, Rene de Cotret S, Lapault S, LeBoeuf D, et al. 2015. *Nature Commun.* 6:6034

[50] Alexandrov AS, Bratkovsky AM. 1996. *Phys. Rev. Lett.* 76:1308–1311

[51] Melikyan A, Vafek O. 2008. *Phys. Rev. B* 78:020502(R)

[52] Pereg-Barnea T, Weber H, Refael G, Franz M. 2010. *Nature Phys.* 6: 44–49

[53] Sebastian SE, Harrison N, Altarawneh MM, Liang R, Bonn DA, Hardy WN, Lonzarich GG. 2010. *Phys. Rev. B* 81:140505

[54] LeBoeuf D, Doiron-Leyraud N, Levallois J, Daou R, Bonnemaison JB, Hussey NE, Balicas L, Ramshaw BJ, Liang R, Bonn DA, Hardy WN, Adachi S, Proust C, Taillefer L. 2007. *Nature* 450:533-536

[55] LeBoeuf D, Doiron-Leyraud N, Vignolle B, Sutherland M, Ramshaw BJ, Levallois J, Daou R, Laliberté F, Cyr-Choinière O, Chang J, Jo YJ, Balicas L, Liang R, Bonn DA, Hardy WN, Proust C, Taillefer, L. 2011. *Phys.Rev. B* 83:054506

[56] Chang J, Daou R, Proust C, LeBoeuf D, Doiron-Leyraud N, Laliberté F, Pingault B, Ramshaw BJ, Liang R, Bonn DA, Hardy WN, Takagi H, Antunes AB, Sheikin I, Behnia K, Taillefer, L. 2010. *Phys. Rev. Lett* 104:057005

[57] Carrington A, Yelland E. 2007. *Phys.Rev. B* 76:140508(R)

[58] Elfimov IS, Sawatzky GA, Damascelli A. 2008. *Phys.Rev. B* 77:060504(R)

[59] Das T. 2012. *Phys.Rev. B* 86:064527

[60] Doiron-Leyraud N, Lepault S, Cyr-Choinière O, Vignolle B, Grissonnanche G, Laliberté F, Chang J, Barišić N, Chan MK, Ji L, Zhao X, Li Y, Greven M, Proust C, Taillefer L. 2013. *Physical Review X* 3:021019

[61] Taillefer L. 2009. *J. Phys.: Condens. Matter* 21:164212

[62] Millis AJ, Norman MR. 2007. *Phys. Rev. B* 76:220503

[63] Chakravarty S, Kee HY. 2008. *Proc. Natl Acad. Sci. USA* 105:8835-8839

[64] Yao H, Lee D-H, Kivelson S. 2011. *Phys. Rev. B* 84:012507

[65] Varma CM. 2009. *Phys. Rev. B* 79:085110

[66] Senthil T, Lee PA. 2009. *Phys. Rev. B* 79:245116

[67] Oh H, Choi HJ, Louie SG, Cohen ML. 2011. *Phys. Rev. B* 84:014518

[68] Chen W-Q, Yang K-Y, Rice TM, Zhang FC. 2008. *Europhys. Lett.* 82:17004

[69] Wu T, Mayaffre H, Kramer S, Horvatic M, Berthier C, Hardy WN, Liang R, Bonn DA, Julien M-H. 2011. *Nature* 477:191-194

[70] Wu T, Mayaffre H, Krämer S, Horvatic M, Berthier C, Kuhns PL, Reyes AP, Liang R, Hardy WN, Bonn DA, and Julien M-H. 2013. *Nature Commun.* 4:2113

[71] Ghiringhelli G, Le Tacon M, Minola M, Blanco-Canosa S, Mazzoli C, Brookes NB, De Luca GM, Frano A, Hawthorn DG, He F, Loew T, Sala MM, Peets DC, Salluzzo M, Schierle E, Sutarto R, Sawatzky GA, Weschke E, Keimer B, Braicovich L. 2012. *Science* 337:821–825

[72] Achkar AJ, Sutarto R, Mao X, He F, Frano A, Blanco-Canosa S, Le Tacon M, Ghiringhelli G, Braicovich L, Minola M, Moretti Sala M, Mazzoli C, Liang R, Bonn DA, Hardy WN, Keimer B, Sawatzky GA, Hawthorn DG. 2013. *Phys. Rev. Lett.* 109:167001

[73] Chang J, Blackburn E, Holmes AT, Christensen NB, Larsen J, Mesot J, Liang R, Bonn DA, Hardy WN, Watenphul A, Zimmermann Mv, Forgan EM, Hayden SM. 2012. *Nat. Phys.* 8:871-877

[74] Blackburn E, Chang J, Hücker M, Holmes AT, Christensen NB, Liang R, Bonn DA, Hardy WN, Rütt U, Gutowski O, Zimmermann Mv, Forgan EM, Hayden SM. 2013. *Phys. Rev. Lett.* 110:137004

[75] LeBoeuf D, Kramer S, Hardy W N , Liang R, Bonn D A, and Proust C. 2013 *Nat. Phys.* 9:79–83

[76] Hinton JP, Koralek JD, Lu YM, Vishwanath A, Orenstein J, Bonn DA, Hardy WN, Liang R. 2013. *Phys. Rev. B* 88:060508(R)

[77] Torchinsky DH, Mahmood F, Bollinger AT, Božovic I, Gedik N. 2013. *Nature Materials*, 12:387–391

[78] Tranquada JM, Sternlieb BJ, Axe JD, Nakamura Y, Uchida S. 1995. *Nature* 375:561–563

[79] da Silva Neto EH, Aynajian P, Frano A, Comin R, Schierle E, Weschke E, Gyenis A, Wen J, Schneeloch J, Xu Z, Ono S, Gu G, Le Tacon M, Yazdani A. 2014. *Science* 343:393

[80] Comin R, Frano A, Yee MM, Yoshida Y, Eisaki H, Schierle E, Weschke E, Sutarto R, He F, Soumyanarayanan A, He Y, Le Tacon M, Elfimov I, Hoffman JE, Sawatzky G, Keimer B, Damascelli A. 2014. *Science* 343:390–392

[81] Tabis W, Li Y, Le Tacon M, Braicovich L, Kreyssig A, Minola M, Dellea G, Weschke E, Veit MJ, Ramazanoglu M, Goldman AI, Schmitt T, Ghiringhelli G, Barišić N, Chan MK, Dorow CJ, Yu G, Zhao X, Keimer B, Greven M. 2014. *Nature Commun.* 5:5875

[82] Croft TP, Lester C, Senn MS, Bombardi A, Hayden SM. 2014. *Phys. Rev. B* 89:224513

[83] Christensen NB, Chang J, Larsen J, Fujita M, Oda M, Ido M, Momono N, Forgan EM, Holmes AT, Mesot J, Huecker M, Zimmermann Mv. 2014. *arXiv:* 1404.3192 (preprint)

[84] Sebastian SE, Harrison N, Balakirev FF, Altarawneh MM, Goddard PA, Liang R, Bonn DA, Hardy WN, Lonzarich GG. 2014. *Nature* 511:61–64

[85] Yamaji K. 1989. *J. Phys. Soc. Jpn* 58:1520–1523

[86] Harrison N, Sebastian SE. 2011. *Phys. Rev. Lett.* 106:226402

[87] Allais A, Chowdhury D, Sachdev S. 2014. *Nature Commun.* 5:5771

[88] Maharaj AV, Hosur P, Raghu S. 2014. Phys. Rev. B 90:125108

[89] Sebastian SE, Harrison N, Altarawneh MM, Liang R, Bonn DA, Hardy WN, Lonzarich GG. 2011. *Nature Commun.* 2:471

[90] Chowdhury, D, Sachdev, S. 2014. *Phys. Rev. B* 90:134516

[91] Meier H, Pépin C, Einenkel M, Efetov KB. 2014. *Phy. Rev. B* 89:195115

[92] Nie L, Tarjus G, and Kivelson SA 2013. *Proc. Nat. Acad. Sci.* doi: 10.1073/pnas.1406019111

[93] Liang R, Bonn DA, Hardy WN. 2006. *Phys. Rev. B* 73:180505

[94] Ramshaw BJ, Sebastian SE, McDonald RD, Day J, Tan BS, Zhu Z, Liang R, Bonn DA, Hardy WN, Harrison N. 2015. *Science* 348:317–320

[95] Loram JW, Mirza KA, Cooper JR, Liang WY. 1993. *Phys. Rev. Lett.* 71:1740–1743

[96] Panagopoulos C, Tallon JL, Rainford BD, Xiang T, Cooper JR, Scott, CA. 2002. *Phys. Rev. B* 66:064501

[97] Loram JW, Mirza KA, Cooper JR, Tallon JL. 1998. *J. Phys. Chem. Solids* 59:2091–94

[98] Yuan HQ, Grosche FM, Deppe M, Geibel C, Sparn G, Steglich F. 2003. *Science* 302:2104–2107

[99] Sebastian SE, Harrison N, Altarawneh MM, Mielke CH, Liang R, Bonn DA, Hardy WN, Lonzarich GG. 2010. *Proc. Natl Acad. Sci. USA* 107:6175-6179

[100] Blanco-Canosa S, Frano A, Schierle E, Porras J, Loew T, Minola M, Bluschke M, Weschke E, Keimer B, Le Tacon M. 2014. *Phys. Rev. B* 90:054513

[101] Huecker M, Christensen NB, Holmes AT, Blackburn E, Forgan EM, *et al.* 2014. *Phys. Rev. B* 90:054514

[102] Sanna S, Allodi G, Concas G, Hillier A D and De Renzi R. 2004 *Phys. Rev. Lett.* 93:207001

[103] Haug D, Hinkov V, Sidis Y, Bourges P, Christensen N B, Ivanov A, Keller T, Lin C T and Keimer B. 2010 *New J. Phys.* 12:105006

[104] Wu T, Mayaffre H, Kramer S, Horvatic M, Berthier C, Lin C T, Haug D, Loew T, Hinkov V, Keimer B, and Julien M H. 2013 *Phys. Rev. B* 88:014511

[105] Efetov KB, Meier H, Pépin C. 2013. *Nat. Phys.* 9:442-446

[106] Sachdev S, La Placa R. 2013. *Phys. Rev. Lett.* 111:027202

[107] Wang, Y, Chubukov AV. 2014. *Phys. Rev. B* 90:035149

[108] Lee, PA. 2014. *Phys. Rev. X* 4:031017

[109] Atkinson WA, Kampf AP, Bulut S. 2015. *New J. Phys.* 17:013025

[110] Thomson A, Sachdev S. 2015. *Phys. Rev. B* 91:115142

[111] Harrison N, Sebastian SE. 2014. *New J. Phys.* 16:063025